\documentstyle[12pt]{article}
\def \ni{\noindent}
\def \be{\begin{equation}}
\def \ee{\end{equation}}
\begin{document}
\oddsidemargin 5mm
\setcounter{page}{0}
\renewcommand{\thefootnote}{\fnsymbol{footnote}}
\newpage
\setcounter{page}{0}
\begin{titlepage}
\begin{flushright}
ISAS/108/91/EP \\
\end{flushright}
\vspace{0.5cm}
\begin{center}
{\Large {\bf Generalised Abelian Chern-Simons Theories \\
             and  their Connection to Conformal Field Theories }} \\
\vspace{1.5cm}
{\bf Marco A. C. KNEIPP
\footnote{E-mail: KNEIPP@ITSSISSA.bitnet or 38028::KNEIPP \\
Work partially supported by CNPq - Brazil }} \\
\vspace{0.8cm}
{\em International School for Advanced Studies (SISSA), Strada Costiera
11, 34014
Trieste, Italy} \\
\end{center}
\vspace{6mm}
\begin{abstract}
We discuss the generalization of Abelian Chern-Simons theories when $\theta
$-angles and magnetic monopoles are included. We map sectors of two
dimensional Conformal Field Theories into these three dimensional theories.
\end{abstract}
\vspace{5mm}
\centerline{July 1991}
\end{titlepage}

\newpage

\renewcommand{\thefootnote}{\arabic{footnote}}
\setcounter{footnote}{0}

Three dimensional Chern-Simons(CS) Gauge Theories are interesting for many
mathematical and physical reasons. As shown by Witten \cite{Witten}, CS
theories can be
used to construct new knot invariants. A mapping between
the three dimensional gauge invariant wave-functions and the conformal
blocks of some Conformal Field Theories(CFT) in two dimensions
[1 - 6] was also estabished.
In terms of physical applications, Abelian CS theory is relevant for the
description of the Fractional Quantum Hall Effect and maybe for High $ T_c $
Supercondutivity \cite{Supercond}. With this last motivation, Iengo and
Lechner \cite{Iengo},
constructed the gauge invariant wave-function of the U(1) CS with a
magnetic monopole and $\theta $-angles on a torus, using the path-integral
approach. A natural question that one can raise is to which CFT this
generalised models are related. The purpose of this letter to consider this
question.
Here, however, we  will adopt the  operatorial
formalism to recover the results in \cite{Iengo} and consider the more
general case of a
multidimensional $ R^d/ \Lambda $ compact Abelian gauge group, where
$ \Lambda $ is a d-dimensional integral lattice.

The Chern-Simons action of a multidimensional $ {\bf R}^d/ \Lambda $ compact
Abelian gauge group with time independent charges is defined as

\be
 S = \frac {1}{4\pi}\left[ \int_{\Sigma \times {\bf R}} {\bf \vec{A}}\wedge
{\bf d\vec{A}} \right]
+ \sum_{\vec{q}}{\vec{q} \vec{A}_{0}(r_q)}
\label{cs-forms}
\end{equation}

\ni where the three dimensional manifold has the structure $\Sigma \times
{\bf R}$, $\Sigma$ being a two dimensional compact Riemann surface
without boundary and ${
\bf R}$  corresponding to the time. The 1-form ${\bf \vec{A} \equiv
A_i } \vec{b}_i $ is a gauge field with $\vec{b}_i $ as basis vector
of a lattice $\Lambda $ with  metric $ K_{ij} \equiv \vec{b}_i
\vec{b}_j $. The static charges $\vec{q} \equiv q_i \vec{b}^*_i$, $q_i \in
{\bf Z} $, belong to the
dual lattice $\Lambda^*$ which is generated by the basis vectors $\vec{b}^*_i$
defined by the relation $\vec{b}^*_i
\vec{b}_j = \delta_{ij}$.

Adopting as boundary condition
the gauge fields going to zero at $t = \pm \infty $ and using complex
coordinates, the action can be rewritten in the form

\be
S = \int_{\Sigma \times {\bf R}}{dx^0dzd\bar z \left[\frac{1}{2\pi}
\vec{A}_{\bar{z}}\partial_0 \vec{A}_z +
\vec{A}_0
\left(  \frac{1}{2\pi} \vec{F}_{z\bar{z}} + \sum_{\vec{q}}{\vec{q}
\delta^2 (z - z_q)}
\right) \right] }
\label{cs-with-charges}
\end{equation}

\ni where we used the fact that the boundary terms vanish due to the
boundary condition at $t = \pm \infty $ and since $\Sigma$ don't have
boundaries.
Here $\vec{A}_0$
appears as Lagrange multiplier with the Gauss law,

\be
\vec{G} \equiv \frac{1}{2 \pi} \vec{F}_{z\bar{z}} + \sum_{\vec{q}}{\vec{q}
\delta^2
(z - z_q)}  ,
\label{gauss-law}
\end{equation}

\ni as a constraint and  $\vec{A}_{\bar z}$
and $\vec{A}_z$ are obviously conjugate variables. It follows that the
components of
$\vec{A}_{\bar z}$ and $\vec{A}_z$ will
satisfy the commutation relation

\be
\left[ \vec{A}_{z}(t,z,\bar z) ,
\vec{A}_{\bar{z}}(t,w,\bar w) \right] =
2\pi i \delta^2 (z-w)
\ee

Choosing the $\vec{A}_0 = 0$ gauge, we obtain an action that is invariant
under time independent gauge transformations which are generated by the Gauss
law. In
order to recover at the quantum level the equation of motion following to
the variation of $A_0$, the physical states, $|\Psi_{ph} >$, must satisfy
$\vec{G}|\Psi_{ph} > = 0$.

The class of allowed gauge transformations depends on the topology of
$\Sigma$. If $\Sigma$ has non trivial cycles, one has discrete large gauge
transformations. We have to establish the action of them on
$|\Psi_{ph} >$. In particular if $\Sigma = T^2$ with modular parameter $\tau =
{\tau}_1 + i{\tau}_2  $, a time independent element of the gauge group
can be written as $g(z,\bar z) = e^{i\vec{\phi}(z,\bar z)\vec{H}}$, where
$\vec{H} \in \Lambda^* $ and the gauge parameter $\vec{\phi} (z,\bar z)$ has
the form

\begin{eqnarray}
\vec{\phi } (z,\bar z) = \vec{\phi }_0 +
2\pi \left[ z \left( \frac{-\bar{\tau} \vec{\phi }_1 + \vec{\phi}_2}
{2i\tau_2}\right)
+ \bar{z} \left( \frac{\tau \vec{\phi }_1 - \vec{\phi_2 }}{2i\tau_2}\right)
\right] +
\hfill  \nonumber  \\
\hfill  \mbox{} + \sum_{\left\{ n_1 ,n_2 \right\} \not= 0}
{\vec{\phi }_{n_1 n_2}
e^{2\pi i\left[ z \left( \frac{-\bar{\tau} n_1 + n_2}{2i\tau_2}\right)
+ \bar{z} \left( \frac{\tau n_1 - n_2}{2i\tau_2}\right) \right]}}
\end{eqnarray}

\ni where $\vec{\phi}_1 , \vec{\phi}_2 \in \Lambda$
to $g$ be single-valued and $\vec{\phi}_0$ are continuos parameters with
$\vec{\phi }_0$ and $\vec{\phi }_0 +
\vec{\lambda} $ being identified for an arbitrary $ \vec{\lambda} \in \Lambda
$ to $g$ be compact. The parameters $\vec{\phi}_0$ and
$\vec{\phi}_{n_1n_2}$ correspond to the small
and $\vec{\phi}_1$ and $\vec{\phi}_2$ to the large
gauge transformations.

We are interested in finding the unitary operator $U(g)$ that produces the
finite time independent
gauge transformations. The small part is obtained by a straightforward
exponentiation. However, if we want an unambiguos expression for all $g$ we
should define $U(g)$ by its properties:

\begin{eqnarray}
U(g)\vec{A_z} U^{\dagger}(g) & = & \vec{A_z} + \partial \vec{\phi }
\label{a-trans} \\
U(g)\vec{A_{\bar z}}U^{\dagger}(g) & = & \vec{A_{\bar{z}}} +
\bar{\partial} \vec{\phi }
\label{abar-trans}
\end{eqnarray}

\ni From these properties we can verify that $U(g)$ generally is a projective
representation having the composition law:

\be
U(g_b)U(g_a) = e^{i\pi \left( \vec{\phi}_{1a} \vec{\phi}_{2b} -
\vec{\phi}_{2a} \vec{\phi}_{1b} \right) }U(g_ag_b)
\end{equation}

\ni Only if $\Lambda $ is an integral lattice, that is
$\vec{v} \vec{u} \in {\bf Z}$ for $\vec{v} , \vec{u} \in \Lambda$,
$U(g)$ will be a faithful representation. One can define a consistent
theory also when the lattice is not integral \cite{PolyII}, however in
the present letter we will limit ourselves to integral lattices.

{}From (\ref{a-trans}) and (\ref{abar-trans}) we obtain that, in a basis which
$\vec{A}_z$ is
diagonalized, $U(g)$ acts on a generic state $\Psi[\vec{A}_z]$ in the
following way:

\be
U(g)\Psi [\vec{A}_z] = e^{- \frac{i}{2\pi } \left[ \int_{\Sigma}{
\frac{1}{2} \partial \vec{\phi} \bar{\partial } \vec{\phi} +
\vec{A}_z\bar{\partial }\vec{\phi} }\right]
- 2\pi i \left( \vec{\theta}_1 \vec{\phi}_1  + \vec{\theta}_2 \vec{\phi}_2
\right) } \Psi [\vec{A}_z +
\partial \vec{\phi } ]
\label{psi-trans}
\end{equation}

\ni where $\vec{\theta_1} $ and $\vec{\theta_2} $ are arbitrary.
The origin of the terms with these parameters
are due the fact that (\ref{a-trans}), (\ref{abar-trans}) and the condition
of a faithful
representation determine the form of $ U(g) $ up to a 1-cocycle. They are
analogous to the $\theta$-angle of QCD.

Due to the fact that $\vec{G}\Psi_{ph}[\vec{A}_z] = 0 $, the physical
states satisfy the relation

\be
 U(g)\Psi_{ph}[\vec{A}_z] = e^{-i \sum_{\vec{q}}\vec{q}
\vec{\phi} (z_q,\bar{z_q})}
\Psi_{ph}[\vec{A}_z]
\label{psi-ph-def}
\end{equation}

\ni It is easily verified that a solution of this equation can be constructed
as

\begin{eqnarray}
\Psi_{ph}[\vec{A}_z]&=&\int{D\vec{\phi} \: e^{i\sum_{\vec{q}} \vec{q}
\vec{\phi}(z_q,
\bar{z_q})} U(g)\Psi[\vec{A}_z]} \nonumber  \\
                    &=&\int{D\vec{\phi} \: e^{i\sum_{\vec{q}} \vec{q}
\vec{\phi}(z_q,
\bar{z_q})}e^{- \frac{i}{2\pi } \left[ \int_{\Sigma}{
\frac{1}{2} \partial \vec{\phi} \bar{\partial } \vec{\phi} +
\vec{A}_z\bar{\partial }\vec{\phi} }\right]
- 2\pi i \left( \vec{\theta}_1 \vec{\phi}_1  + \vec{\theta}_2 \vec{\phi}_2
\right) }
 \Psi [\vec{A}_z + \partial \vec{\phi } ] }
\label{psi-ph-general}
\end{eqnarray}

\ni where $\Psi[\vec{A}_z ]$ is a arbitrary state. Here we see
that $\Psi_{ph}[\vec{A}_z]$ corresponds to the correlator of the vertex
operators $e^{i \vec{q} \vec{\phi}(z_q,\bar{z_q})} $ of a free scalar CFT
on $\Sigma$, compactified on the lattice $\Lambda $ and with an external gauge
field $\vec{A}_z$.

Comparing (\ref{psi-trans}) and (\ref{psi-ph-def})we see that the physical
states fulfil the relation:

\be
\Psi_{ph}[\vec{A}_z + \partial \vec{\phi}] = e^{ \frac{i}{2\pi } \left[
\int_{\Sigma}{
\frac{1}{2} \partial \vec{\phi} \bar{\partial } \vec{\phi} +
\vec{A}_z \bar{\partial }\vec{\phi} }\right]
+ 2\pi i \left( \vec{\theta}_1 \vec{\phi}_1 + \vec{\theta}_2
\vec{\phi}_2  \right)
- i \sum_{\vec q}{\vec{q} \vec{\phi} (z_q,\bar{z_q})}}
\Psi_{ph}{\left[\vec{A}_z  \right] }
\label{psi-ph-const}
\end{equation}

\ni Using an arbitrary constant gauge transformation, $\vec{\phi }
(z,\bar z ) = \vec{\phi }_0$, in the last relation we obtain the
condition that $\sum_{\vec{q}} \vec{q } = 0$ for the physical states.

On the torus, the $\vec{A}_z$ component can be decomposed in the form

\begin{eqnarray}
\vec{A}_z( z , \bar z ) & = &\frac{i\pi}{\tau_2}\vec{a} +
\sum_{\left\{ n_1 ,n_2 \right\} \not= 0}{\vec{a}_{n_1 n_2}
e^{2\pi i\left[ z \left( \frac{-\bar{\tau} n_1 + n_2}{2i\tau_2}\right)
+ \bar{z} \left( \frac{\tau n_1 - n_2}{2i\tau_2}\right) \right]}}
\nonumber \\
                & = &\frac{i\pi }{\tau_2 } \vec{a} + \partial
\vec{\phi}_{sm}
\end{eqnarray}

\ni where the last term can be eliminated by a small gauge transformation
$\vec{\phi}_{sm} $. Therefore, to calculate a generic physical state $\Psi_{ph}
[\vec{A}_z] $, it
is enough to find $\Psi_{ph} \left( \frac{i\pi }{\tau_2 } \vec{a} \right) $ and
then, from (\ref{psi-ph-const}), we arrive

\be
\Psi_{ph} \left[ \vec{A}_z(z,\bar z)\right] = e^{ \frac{i}{2\pi } \left[
\int_{\Sigma }{\frac{1}{2} \partial \vec{\phi}_{sm} \bar{\partial }
\vec{\phi}_{sm} }\right]
- i \sum_{\vec q}{\vec{q} \vec{\phi}_{sm} (z_q,\bar{z_q} )}}
\Psi_{ph} \left( \frac{i\pi }{\tau_2 } \vec{a} \right)
\end{equation}

\ni To determine the form of $ \Psi_{ph} \left( \frac{i\pi }{\tau_2 }
\vec{a} \right) $, we use another time (\ref{psi-ph-const}) and obtain the
quasi-periodicity relation:

\begin{eqnarray}
\lefteqn{\Psi_{ph} \left( \frac{i\pi }{\tau_2 } \left(\vec{a} -
\vec{\phi_2 } + \bar{\tau }
\vec{\phi_1 } \right) \right) = }   \label{psi-ph-period}   \\
& &  = e^{\frac{\pi}{\tau_2 } \left[ \vec{a} \left( \tau \vec{\phi}_1 -
\vec{\phi}_2
\right) + \frac{1}{2} \left( \tau \vec{\phi}_1 - \vec{\phi}_2 \right)
\left( \bar{\tau}
\vec{\phi}_1 - \vec{\phi}_2 \right) \right]
+ 2 \pi i \left( \vec{\theta}_1 \vec{\phi}_1
+ \vec{\theta}_2 \vec{\phi}_2 \right)
+i\sum_{\vec{q}} {2\pi \vec{q}
\left[ z \left( \frac{-\bar{\tau} \vec{\phi }_1 + \vec{\phi}_2}
{2i\tau_2}\right)
+ \bar{z} \left( \frac{\tau \vec{\phi }_1 - \vec{\phi_2 }}{2i\tau_2}\right)
\right]}}
\Psi_{ph} \left( \frac{i\pi }{\tau_2 } \vec{a} \right)
\nonumber
\end{eqnarray}

\ni For the case without charges, this relation has the independent
(non-normalized) solutions

\begin{equation}
\Psi_{\vec{\beta} } \left( \frac{i\pi\vec{a}}{\tau_2 } \right) =
\exp{ \left( \frac{\pi \vec{a}^2}{2\tau_2} \right)}
\sum_{\vec{\alpha} \in \Lambda }{ e^{-2\pi i \vec{\theta}_1 \vec{\alpha}}
\exp \left[ -i\pi \bar{\tau}
\left( \vec{\alpha}  + \vec{\beta}  + \vec{\theta}_2 \right)^2
- 2\pi i \left( \vec{\alpha}  + \vec{\beta}  + \vec{\theta}_2 \right)
\vec{a} \right] }
\label{psi-ph-theta}
\end{equation}

\ni where $\vec{\beta}  \in  \Lambda^* / \Lambda $ .

Since in the $A_0 = 0$ gauge the action is invariant under modular
transformations, we have to guarantee that these physical states form a
representation of the modular transformations

\be
\begin{array}{cc}
T : \left\{
    \begin{array}{l}
    x_1' = x_1 + x_2 \\
    x_2' = x_2
    \end{array} \right.  &
S : \left\{
    \begin{array}{l}
    x_1' = x_2 \\
    x_2' = - x_1
    \end{array} \right.
\end{array}
\ee

\ni The unitary operators which implement these modular transformations are
given
by \cite{Verlinde}

\be
T = \eta \exp \frac{i}{4\pi} \vec{a}_1^2 \hspace{7mm}
S = \eta ' \exp \frac{i}{8\pi} \left[\vec{a}_1^2 + \vec{a_2^2} \right]
\ee

\ni where $ \eta $ and $\eta '$ are some phases and $ \vec{a}_1 +
\bar{\tau}\vec{a}_2 = \vec{a}$. From the T transformation we obtain the
condition that

\begin{equation}
\frac{\vec{\lambda} \vec{\lambda} }{2} + \vec{\lambda} \vec{\theta}_2 \in
{\bf Z} \hspace{10mm} \forall \vec{\lambda} \in \Lambda
\label{t-condition}
\end{equation}

\ni To solve it, we use the fact that for an arbitrary $\Lambda $ integral,
the vector basis can be decomposed as
$ \vec{b_i } = \left\{ \vec{o_1 }, \ldots  ,\vec{o_n } , \vec{e }_{n+1 } ,
\ldots , \vec{e_d } \right\} $ with

\begin{eqnarray}
\vec{e_i }  \vec{e_i } & \in &  2{\bf Z}  \\
\vec{o_i }  \vec{o_i } & \in &  2{\bf Z} + 1
\end{eqnarray}

\ni We verify that the general solution of (\ref{t-condition}) is

\begin{equation}
\vec{\theta}_2 = \sum_j{\frac{\vec{o_j}^*}{2}}  \pmod{\Lambda^*}
\end{equation}

\ni The covariance of the wave-function under S transformation
impose that

\be
\vec{\theta}_1 = \vec{\theta}_2 \pmod{\Lambda^*}
\ee

For the case of a even lattice (that is, a lattice generated by a basis
composed only of vectors with even norm), we will have that $\vec{\theta}_1 =
\vec{\theta}_2 = 0$ and we recover the results in \cite{Moore,Bos}. However,
the important
consequence of considering the $\theta$-angles is that it opens the
possibility to consider a ${\bf R}^d/ \Lambda$ CS with an arbitrary
$\Lambda$. We don't have any more the constraint that $\Lambda$ is
even as normally considered.

Now we will see that the physical wave-functions (\ref{psi-ph-theta})
are the conformal blocks of sectors of two differents CFT's defined on
$T^2$. The first one is a CFT defined by a chiral algebra generated by
$G_i\equiv z^{\vec{b}^2_i/2} :e^{\left(i \vec{b_i} \vec{\phi}(\bar z )\right)}:
$ and
$ \vec{G}_0 \equiv i\partial_{\bar{z}} \vec{\phi}(\bar z )$ with the chiral
scalar field $\vec{\phi}(\bar{z}) = \vec{q} - i\vec{p} \ln \bar{z} + i\sum_{n
\not= 0}{\vec{\alpha}_n\bar{z}^n} $. The generators $G_i$ have scaling
dimensions $\vec{b_i}^2 /2$ that can be integer or half-integer and we will
call them as "bosonic" and "fermionic"
respectivaly. The fermionic set has two more features with respect to the
bosonic one. The first feature is that the chiral algebra is compatible which
Ramond(periodic) or Neveu-Schwarz(anti-periodic) boundary conditions. The
requirement that the $G_i$'s are periodic, selects for the Ramond sector
the momenta $ \vec{\lambda}^* + \sum_i{\vec{o}_i^* /2} $  and
for the the Neveu-Schwarz sector the momenta
to $\vec{\lambda}^* $, $\forall \vec{\lambda}^* \in \Lambda^* $. The second
feature comes from the fact that since the fermionic
operators have semi-integer dimensions, they would produce monodromies
inside the representations. Therefore we decompose the representations
through a projection operator. This projection should separate the states
connected by a even number of applications of the fermionic operators.
Let's consider a generic state $|\vec{\alpha} + \vec{\beta}>$ produced from
the highest weight state $|\vec{\beta}>$ with $\vec{\alpha} =
\alpha^{odd}_i \vec{o}_i + \alpha^{even}_i \vec{e}_i$. We can verify that the
operators $ \frac{1 \pm (-1)^F}{2} $ with $F \equiv \sum_i{\alpha_i^{odd}} =
2\vec{\alpha} \vec{\theta}$, produces the wanted projection.

It is known that modular transfomations generaly mix the
different boundaries conditions and only the Ramond-Ramond sector by itself
is invariant under modular transformations. To obtain the
corresponding characters of
the Ramond-Ramond sector for our chiral algebra, we proceed exactly in
the same way as for the ordinary
fermionic case: the path integral with periodic boundary
conditions in both directions corresponds to the trace of the operator
$(-1)^F \exp(-2\pi i \bar{\tau}L_0 - \vec{a}\vec{G_0} )$
over the states in a representation
with momenta  $ \vec{\lambda}^* + \sum_i{\vec{o}^*_i /2} $.
This will give as
the result for a representation built from
a highest weight state $|\vec{\beta}>$, $ \beta \in \Lambda^*/
\Lambda$

\begin{equation}
\chi_{\vec{\beta}} = \sum_{\vec{\alpha} \in \Lambda }{ \left( -1 \right)^F
\exp \left[ -i\pi \bar{\tau}
\left( \vec{\alpha}  + \vec{\beta}  + \sum_i{\vec{o}^*_i /2} \right)^2
- 2\pi i \left( \vec{\alpha}  + \vec{\beta}  + \sum_i{\vec{o}^*_i /2} \right)
\vec{a} \right] }
\end{equation}

\ni which corresponds to difference of the conformal blocks that comes from
the sectors $ \frac{1 + (-1)^F}{2} $ and $ \frac{1 - (-1)^F}{2} $ with
momenta to $ \vec{\lambda}^* + \sum_i{\vec{o}^*_i /2} $.
Comparing this
last result with the wave-functions (\ref{psi-ph-theta}) and using our
definition of F, we
can conclude that our physical wave-funtions are the characters of the
Ramond-Ramond sector of the above CFT.

The second CFT is defined from a even lattice that we will
denote by $\tilde{\Lambda}$. In this lattice we substitute the odd vector basis
$\vec{o}_i$ of our original $\Lambda$ by $\vec{f}_i \equiv 2 \vec{o}_i $ and
leave unchanged the even vector basis $\vec{e}_i$. Correspondly, in the new
dual lattice, $\tilde{\Lambda}^*$, we substitute $\vec{o}_i^*$ by
$\vec{f}_i^* \equiv \vec{o}_i^* / 2$ in order to preserve the relation $
\vec{b}_i \vec{b}_j^* = \delta_{ij}$, $ \vec{b_i } = \left\{ \vec{f_1 },
\ldots  ,\vec{f_n } , \vec{e }_{n+1 } ,
\ldots , \vec{e_d } \right\} $. This is a special kind of even
lattice since more than the condition $\vec{b}_i \vec{b}_i \in 2{\bf Z}$,
we have that also $\vec{b}_i \vec{f}_j \in 2{\bf Z}$.

{}From this lattice we define as before a chiral algebra, but now containig
only bosonic operators. We can verify that for this lattice we have the
discrete global symmetry property that the $G_i$'s that are connected by
shifts of $
\vec{f}_j /2$, have the same scaling dimension (modulo an integer).
It is not difficult to prove that the following combination of conformal
blocks form a representation of modular group:

\be
\chi_{\vec{\beta}} = \tilde{\chi}_{\vec{\beta}} -
\sum_{i = 1}^n{\tilde{\chi}_{\vec{\beta} + \frac{1}{2}\vec{f}_i}} +
\sum_{i = 1}^j{\sum_{j = 2}^n {\tilde{\chi}_{\vec{\beta} + \frac{1}{2}\left(
\vec{f}_i + \vec{f}_j \right)}}} - \cdots
+ (- 1)^n \tilde{\chi}_{\vec{\beta} + \frac{1}{2}\left(\vec{f}_1 + \vec{f}_2
+ \cdots + \vec{f}_n \right)}
\ee

\ni where

\begin{eqnarray*}
\tilde{\chi}_{\vec{\gamma}} & \equiv & \sum_{\vec{\alpha} \in \tilde{\Lambda}}
{\exp \left[ -i\pi\bar{\tau} \left(\vec{\alpha} + \vec{\gamma}\right)^2 -
2\pi i \left(\vec{\alpha} + \vec{\gamma}\right) \vec{a} \right] } \\
\vec{\beta} & \in & \sum_{m_i,n_i \in {\bf Z}}{ \left[ \left(2m_i + 1 \right)
\vec{f_i}^* + n_i\vec{e_i}^* \right] } / \sum_{p_i,q_i \in {\bf Z}}{\left[
\frac{p_i}{2}\vec{f}_i + q_i\vec{e}_i \right]}
\end{eqnarray*}

\ni We can put it in a compact form:

\be
\chi_{\vec{\beta}} = \sum_{\vec{g}}{(-1)^{2\vec{g}\sum_i
{\vec{f_i}^*}} \tilde{\chi}_{\vec{\beta} + \vec{g}} }
\ee

\ni where $\vec{g}$ are the different shift produced by the combination of
$\vec{f}_i/2$. If we rewrite the wave-function (\ref{psi-ph-theta}) using the
vector basis of $\tilde{\Lambda}$, we obtain the same modular matrices.

We discuss now the quantisation of our theory in the presence of a magnetic
monopole.
Following the procedure of t'Hooft \cite{t'Hooft}, we can construct a
configuration
with magnetic monopole on $T^2$ by using the boundary conditions:

\begin{eqnarray}
\vec{A}_i (x^1,1) = \vec{A}_i(x^1,0) + \partial_i \vec{r}(x^1,0) \hspace{5mm}
i=1,2  \nonumber \\
\vec{A}_i (1,x^2) = \vec{A}_i(0,x^2) + \partial_i \vec{s}(0,x^2) \hspace{5mm}
i=1,2 \\
\vec{s}(0,1) - \vec{s}(0,0) - \vec{r}(1,0) + \vec{r}(0,0) \in 2\pi \Lambda
\nonumber
\end{eqnarray}

\ni where the last condition guarantee that the Dirac string
stay invisible. Therefore, we can consider that $\vec{r}  = 2\pi \vec{\mu}x^1
$ and $\vec{s} = 0 $ where $\vec{\mu} \in \Lambda$.
Passing to complex coordinates, the gauge fields and the field strength

\begin{eqnarray}
\vec{A}_z & = & \vec{A}_z^p + \frac{\bar{\tau} \pi}{2\tau_2^2 }
\left( z - \bar z
\right) \vec{\mu}
\nonumber  \\
\vec{A}_{\bar z} & = & \vec{A}_{\bar{z} }^p - \frac{\tau \pi}{2\tau_2^2 }
\left(  z - \bar z \right) \vec{\mu}  \label{monopole} \\
\vec{F}_{z\bar{z} }  & = & \vec{F}_{z\bar{z} }^p + \frac{\pi }{i\tau_2 }
\vec{\mu }
\nonumber
\end{eqnarray}

\ni fulfil the above boundary conditions, where $\vec{A}_z^p ,
\vec{A}_{\bar{z}}^p
$ and $\vec{F}^p_{z\bar{z}} $ correspond to the field
configuration with periodic boundary conditions. Substituting
(\ref{monopole}) in (\ref{cs-forms}),
and doing some integration by parts, we obtain

\be
S = \int_{\Sigma \times {\bf R}}{dx^0dzd\bar z \left[\frac{1}{2\pi}
\vec{A}_{\bar{z}}^p \partial_0 \vec{A}_z^p +
\vec{A}_0
\left(  \frac{1}{2\pi} \vec{F}_{z\bar{z}}^p + \sum_{\vec{q}}{\vec{q}  \delta^2
(z - z_q)} + \frac{1 }{2 i\tau_2 }
\vec{\mu } \right)\right]  }
\label{cs-with-monopole}
\end{equation}

\ni plus some non periodic terms. Using as guiding principle to consider only
truely periodic quantities in the action, we will consider
(\ref{cs-with-monopole}) as a new definition
of the action, replacing (\ref{cs-forms}). This action has a vanishing
hamiltonian as
the previous one. Comparing with (\ref{cs-with-charges}), we see that the only
effect of the
inclusion of a magnetic monopole was a change in the Gauss law. Therefore, the
physical states will satisfy the new condition:

\be
 U(g)\Psi_{ph}[\vec{A}_z] = e^{-i \sum_{\vec{q}} \vec{q}
\vec{\phi} (z_q,\bar{z_q})
+ i\vec{\mu} \vec{\phi_0}} \Psi_{ph}[\vec{A}_z]
\label{psi-ph-monopole-def}
\end{equation}

\ni Here we put only the constant part of $\vec{\phi}$ for the magnetic
monopole, since the part that
comes from the large gauge transformation can be absorved by a redefinition of
the $\theta$-angles. A general solution to this equation will be:

\begin{eqnarray}
\Psi_{ph}[\vec{A}_z] &=& \int{D\vec{\phi} \: e^{i\sum_{\vec{q}} \vec{q}
\vec{\phi}(z_q,
\bar{z_q}) - i\vec{\mu} \vec{\phi_0}}
U(g)\Psi[\vec{A}_z]} \nonumber \\
                     &=& \int{D\vec{\phi} \: e^{i\sum_{\vec{q}} \vec{q}
\vec{\phi}(z_q,
\bar{z_q}) - i\vec{\mu} \vec{\phi_0}}
e^{- \frac{i}{2\pi } \left[ \int_{\Sigma}{
\frac{1}{2} \partial \vec{\phi} \bar{\partial } \vec{\phi} +
\vec{A}_z\bar{\partial }\vec{\phi} }\right]
- 2\pi i \left( \vec{\theta}_1 \vec{\phi}_1  + \vec{\theta}_2 \vec{\phi}_2
\right) } \Psi [\vec{A}_z +
\partial \vec{\phi } ] }
\label{psi-ph-monopole-general}
\end{eqnarray}

\ni From (\ref{psi-ph-monopole-def}) and (\ref{psi-trans}) we arrive that
$\Psi_{ph}$ will by constrained by:

\begin{equation}
\Psi_{ph}[\vec{A}_z + \partial \phi ] = e^{\frac{i}{2\pi } \left[
\int_{\Sigma}{
\frac{1}{2} \partial \vec{\phi} \bar{\partial } \vec{\phi} +
\vec{A}_z\bar{\partial }\vec{\phi} }\right]
+ 2\pi i \left( \vec{ \theta}_1 \vec{\phi}_1  + \vec{\theta}_2 \vec{\phi}_2
\right)
- i \sum_{\vec q}{\vec{q} \vec{\phi} (z_q,\bar{z_q})} + i\vec{\mu}
\vec{\phi}_0 }
\Psi_{ph}{\left[ \vec{A}_z \right] }
\label{psi-ph-monopole-const}
\end{equation}

\ni Using a constant gauge transformation we obtain the new condition
$\sum_{\vec{q}}{\vec{q}} = \vec{\mu} $ for the
physical states. Therefore, when there is a magnetic monopole the charge
distribution must have a total charge different from zero. Moreover,
since $\vec{q} \in \Lambda^*$ and $\vec{\mu} \in \Lambda$ this relation
impose that $\Lambda \subset \Lambda^*$, which means that in the presence of a
magnetic monopole $\Lambda$ is necessarily an integral lattice.

{}From (\ref{psi-ph-monopole-const}) we arrive to the same
quasi-periodicity relation (\ref{psi-ph-period}) but with the condition
$\sum_{\vec{q}}{\vec{q}} = \vec{\mu} $. This relation will have the solutions

\begin{equation}
\Psi_{\vec{\beta} } \left( \frac{i\pi\vec{\tilde{a}}}{\tau_2 } \right) =
\exp{ \left( \frac{\pi \vec{\tilde{a}}^2}{2\tau_2} \right)}
\sum_{\vec{\alpha} \in \Lambda }{ (-1)^F
\exp \left[ -i\pi \bar{\tau}
\left( \vec{\alpha}  + \vec{\beta}  + \vec{\theta} \right)^2
- 2\pi i \left( \vec{\alpha}  + \vec{\beta}  + \vec{\theta} \right)
\left( \vec{a - \sum_{\vec{q}}{\vec{q}\bar{z}_q} } \right) \right] }
\end{equation}

\ni where $\vec{\beta} \in \Lambda^* /\Lambda $ and $\vec{\tilde{a}} \equiv
\vec{a} + \sum_{\vec{q}}{ \vec{q} \left( z_q - \bar{z}_q \right)}$. Therefore,
for the one dimensional lattice, we recover the results in \cite{Iengo}. We
can conclute that the only consequence of the inclusion of the monopole is
that now $\sum_{\vec{q}}{\vec{q}} = \vec{\mu} $.

The condition of a total charge different from zero appears already at the
classical level: as $\vec{q}$ couples directly to the magnetic field in the
equations of motion (\ref{gauss-law}), the introduction of a magnetic monopole
changes
the charge conservation law from $\sum_{\vec{q}}{\vec{q}} = 0$ to
$\sum_{\vec{q}}{\vec{q}} -
\vec{\mu} = 0 $. This result can be extented to other Riemann surfaces.
In particular we can do the same for the sphere. Therefore we can recognize
the condition $ \sum_{\vec{q}} \vec{q} \not= 0 $ as the same
considered by Dotsenko and Fateev \cite{Dotsenko} for the Feigin-Fuchs
construction on
the sphere. There, through the introduction of a term $\int{ R \sqrt{g} \phi}$
in the action and using after complex coordinates, they arrive in a correlator
similar to (\ref{psi-ph-monopole-general}).
It seems therefore that there is exist a connection between the CS
theory with a magnetic monopole and minimal models.  However, the screening
operators necessary for the Feigin-Fuchs construction don't have a
clear  interpretation from the CS point of view.

\vspace{10mm}

\ni {\Large{\bf Acknowledgements}}

I would like to thank A. Schwimmer for the suggestion of this problem and
for the enlightening and patient conversations.

\end{document}